\documentclass[%
 reprint,
superscriptaddress,
 amsmath,amssymb,
 aps,
]{revtex4-2}

\usepackage{graphicx}
\usepackage{caption}
\usepackage{subcaption}
\usepackage{dcolumn}
\usepackage{bm}
\usepackage{siunitx}
\usepackage{hyperref} 
\usepackage{cleveref}
\crefname{figure}{Fig.}{Figs.}
\Crefname{figure}{Fig.}{Figs.}

\begin{document}

\preprint{APS/123-QED}

\title{Tunable Terahertz Bandpass Filter based on the Photonic Emulation of a 2DEG Interferometer through the Geometric Phase of Light}

\author{Victor I. Iwanaga}
\affiliation{Department of Physics, Graduate School of Science, Tohoku University, Sendai 980-8578, Japan}

\author{Hiroaki Shirasaka}
\affiliation{Department of Physics, Graduate School of Science, Tohoku University, Sendai 980-8578, Japan}

\author{Shin'ichiro Hayashi}
\affiliation{National Institute of Information and Communications Technology, Tokyo 184-8795, Japan}

\author{Katsuhiko Miyamoto}
\affiliation{Graduate School of Engineering, Chiba University, Chiba 263-8522, Japan}
\affiliation{Molecular Chirality Research Center, Chiba University, Chiba 263-8522, Japan}

\author{Yoshiharu Urata}
\affiliation{PHLUXi, Inc., Sendai 981-0952, Japan}

\author{Kouji Nawata}
\affiliation{Department of Information and Communication Engineering, Tohoku Institute of Technology, Sendai 982-8577, Japan}

\author{Jun-ichi Shikata}
\affiliation{College of Engineering, Nihon University, Koriyama 963-8642, Japan}

\author{Norihiko Sekine}
\affiliation{National Institute of Information and Communications Technology, Tokyo 184-8795, Japan}

\author{Seigo Ohno}
\email[Corresponding author: ]{seigo@tohoku.ac.jp}
\affiliation{Department of Physics, Graduate School of Science, Tohoku University, Sendai 980-8578, Japan}


\begin{abstract}
Terahertz frequencies are expected to become a key player in next-generation wireless communications.
For this transition to occur seamlessly, tunable bandpass filters for channel selection are essential. However, conventional Fabry–Pérot cavity-based bandpass filters face inherent limitations. Specifically, their tunability is restricted by physical constraints, such as the maximum achievable mirror separation.
In this work, drawing inspiration from electronic interferometers, we demonstrate a continuously tunable bandpass filter operating near 0.3 THz by exploiting the geometric phase of light, the Pancharatnam–Berry phase. Our device incorporates intracavity waveplates for polarization control and is described by a mathematical expression analogous to that of the electronic interferometer governed by the Aharonov–Bohm phase, thereby realizing a photonic analogue.
Using custom-designed subwavelength waveplates and cavity mirrors, we achieve a finesse of $\mathcal{F}\approx17$, a quality factor of $Q\approx1.5\times10^{3}$, and a resonance linewidth of $\Delta f_{\mathrm{FWHM}}\approx\qty{200}{\mega\hertz}$.
Beyond terahertz communication technology, this system is expected to find applications in fields requiring high frequency resolution, such as spectroscopy.

\end{abstract}

\maketitle


\section{\label{sec:level1}Introduction}
Different branches of physics have long driven the development of one another through the exchange of ideas and concepts previously believed to be unique to specific fields.
For example, the periodic nature of atomic crystals leading to electronic bandgaps between the valence and conduction bands motivated the development of photonic crystals, where periodic dielectric structures give rise to photonic bandgaps \cite{Yablonovitch, Joannopoulos2008}.

Similarly, since its initial discovery by Pancharatnam \cite{Pancharatnam1956} and later generalization by Berry \cite{Berry01111987} to quantum systems, the concept of geometric phases has been recognized across a wide range of physical systems, establishing connections between different branches of physics \cite{Tomita, interf, Cohen2019GeometricPhase}.
In photonic systems, geometric and topological concepts have been explored through analogies with electronic phenomena \cite{photonic_spin_hall, topological_photonics}. 
For example, synthetic magnetic fields for light have been experimentally demonstrated, enabling the realization of photonic Landau levels \cite{Schine2016, Pan2025}.
These developments provide new degrees of freedom for structuring light fields through geometric and topological effects, while enabling analogies between photonic and electronic systems.

Unlike the dynamical phase, the geometric phase depends on the path traversed by a system in parameter space rather than the duration of its evolution.
For a closed path, the acquired geometric phase is related to the curvature enclosed by the path.
A prominent example is the Aharonov–Bohm (AB) phase \cite{AB_phase}, where an electron traversing a closed path acquires a phase determined by the enclosed magnetic flux.
In a two-dimensional electron gas (2DEG) configured as a Fabry–Pérot (FP) interferometer, the AB phase can be observed through interference, with the conductance oscillating periodically as the enclosed magnetic flux is varied \cite{setup_pioneer, AB_tuning_main}.
In optics, on the other hand, a geometric phase known as the Pancharatnam–Berry (PB) phase arises when the polarization state of light evolves along a closed path on the Poincaré sphere.
Although the underlying physical systems are fundamentally different, both the AB and PB phases depend on global properties of closed paths, namely the enclosed magnetic flux and the solid angle subtended on the Poincaré sphere, respectively, rather than on the local dynamical evolution along the path.

To explore this analogy, we develop an optical FP cavity in which waveplates introduce a PB phase that tunes the cavity modes, analogous to the tuning of electronic interferometer modes by magnetic flux via the AB phase.
While this concept, known as a \textit{Spectral Drill cavity }\cite{Ohno:18}, was previously demonstrated experimentally at a wavelength of \qty{1.55}{\micro\meter}, no explicit parallels to electronic interferometers were established.
In this work, we explicitly show the shared mathematical structure between the two systems, offer greater experimental freedom, and implement the concept for the first time in the terahertz (THz) regime at \qty{0.3}{\tera\hertz}.

While the PB phase has been widely used in THz metasurfaces \cite{deng2021functional}, and cavity eigenmode control in other frequency regimes has been demonstrated using cavity mirror geometry \cite{Uphoff_2015} or intracavity spatially inhomogeneous waveplates \cite{naidoo2015controlled, Maguid2018, Huang:18, Chen}, its application to control cavity eigenmodes in the THz regime remains unexplored. 
Here, we demonstrate the first implementation of the PB phase in a THz cavity for eigenmode control through rotation of intracavity waveplates.

Substantial efforts are being made to develop electronic devices, sources, and detectors operating at sub-THz frequency of \qty{0.3}{\tera\hertz} for next-generation wireless communications \cite{jiang2024terahertzcommunicationssensing6g}.
Achieving a seamless transition to the \qty{0.3}{\tera\hertz} band requires compact, tunable bandpass filters to isolate and separate individual communication channels.
Such components must deliver precise frequency selectivity and effective noise suppression while maintaining low signal loss.
By providing dynamic control over cavity eigenmodes in the THz regime, we propose the THz Spectral Drill cavity for such bandpass filtering applications

The remainder of this paper is organized as follows. \Cref{sec:concept} introduces the theoretical analogy between the 2DEG interferometer and the Spectral Drill cavity. \Cref{sec:exp_setup} details the design of custom THz optical components required for the constructed Spectral Drill cavity. \Cref{sec:results} presents the experimental characterization, including the achieved frequency resolution for bandpass filter applications, followed by concluding remarks in \cref{sec:conclusion}.

\section{\label{sec:concept}Concept}

\begin{figure*}[t!]
    \centering

    \begin{minipage}[t]{0.55\textwidth}
        \makebox[\textwidth][l]{(a)}
        \vspace{-2mm}
        \centering
        \includegraphics[width=\textwidth]{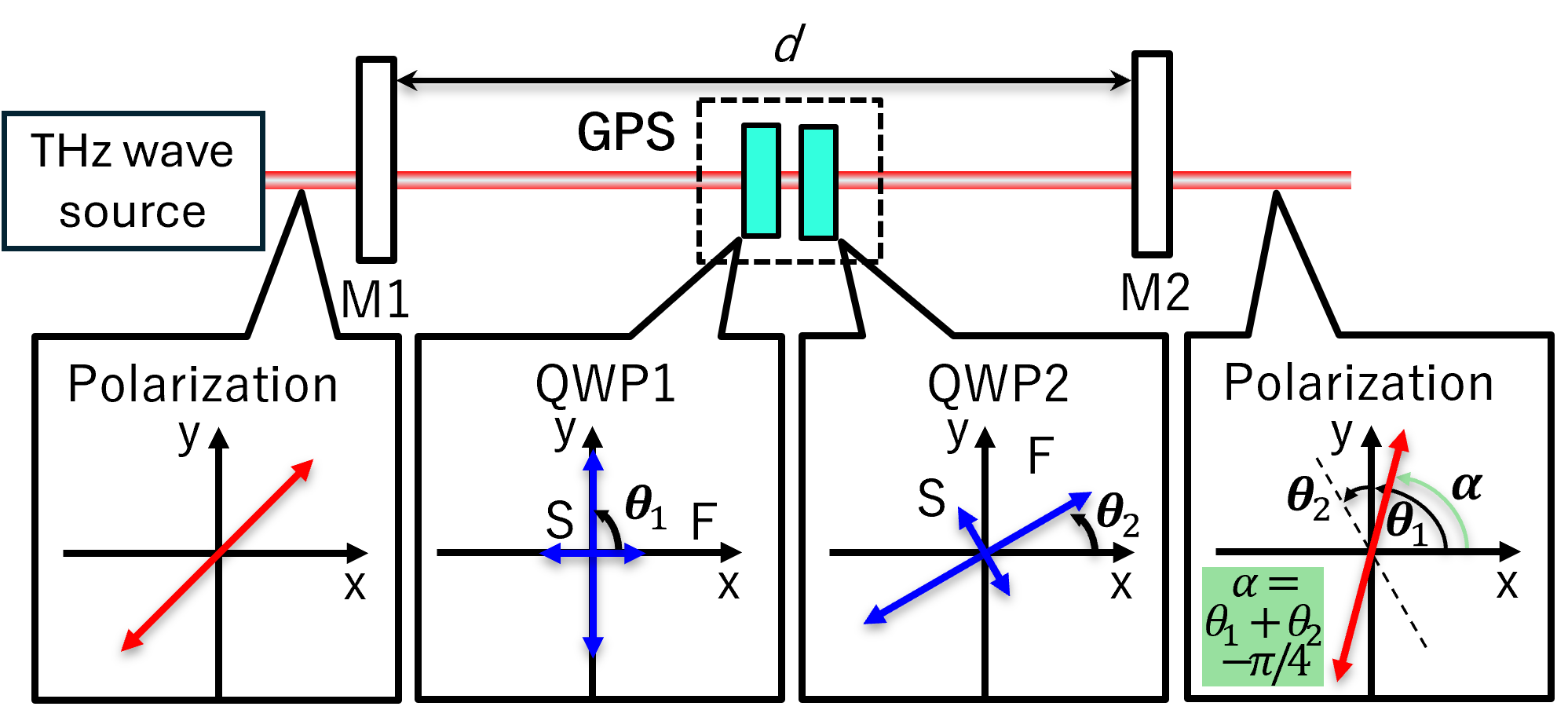}
        \label{fig:drill_schematic}
    \end{minipage}
    \hspace{-3mm}
    \begin{minipage}[t]{0.32\textwidth}
        \makebox[\textwidth][l]{(b)}
        \vspace{-2mm}
        \centering
        \includegraphics[
            width=\textwidth,
            trim=5mm 2mm 0mm 9mm,   
            clip
        ]{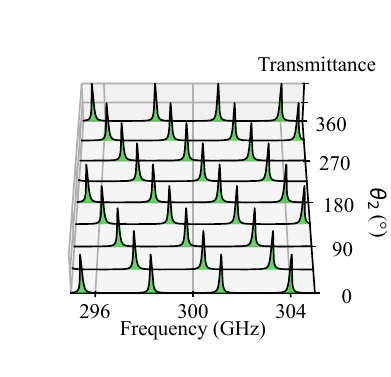}
        \label{fig:3D_drill}
    \end{minipage}

    \vspace{-4mm}

    \begin{minipage}[t]{0.87\textwidth}
        \makebox[\textwidth][l]{(c)}
        \vspace{-3mm}
        \centering
        \includegraphics[
            width=\textwidth
        ]{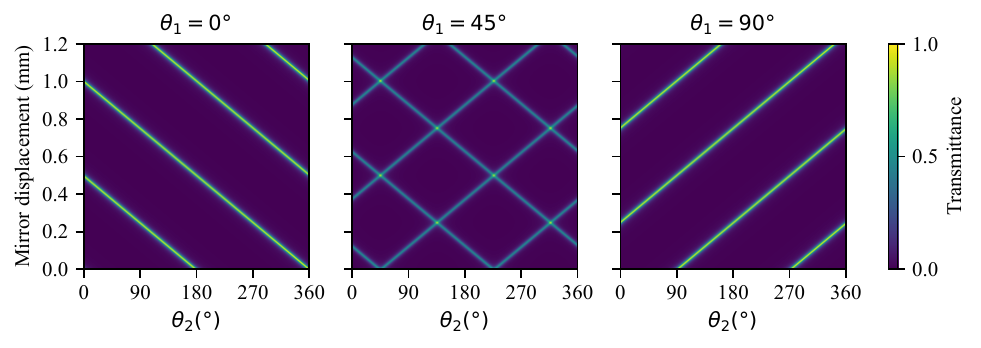}
    \end{minipage}
    
    \caption{Concept of the Spectral Drill cavity: (a) Schematic illustration of the Spectral Drill cavity consisting of a geometric phase shifter (GPS) composed of a fixed QWP (QWP1) and rotatable QWP (QWP2) inserted into a FP cavity.
    The angle $\theta_1$ denotes the orientation of QWP1 relative to a $45^\circ$ offset from the linear input polarization, and $\theta_2$ is the rotation angle of QWP2.
    An incident THz beam undergoes multiple reflections inside the cavity and acquires a PB phase shift with each roundtrip.
    (b) Analytically calculated cavity transmittance as a function of incident frequency and QWP2 rotation angle $\theta_\text{2}$ for QWP1 angle $\theta_\text{1}=\ang{0}$ , where the additional PB phase continuously shifts the axial modes, returning to the initial configuration every \ang{180}.
    (c) Analytically calculated cavity transmittance as a function of mirror displacement and QWP2 rotation angle $\theta_{\text{2}}$ at \qty{0.3}{\tera\hertz} for QWP1 orientation angles $\theta_{\text{1}}$ of \ang{0}, \ang{45}, and \ang{90}. For (b) and (c), a mirror reflectance of $R=0.90$ was assumed.}
    \label{fig:wide}
\end{figure*}

In a 2DEG configured as an electronic FP interferometer, two quantum point contacts (QPCs) act as partially transmitting electronic mirrors, forming a cavity. Under a strong perpendicular magnetic field, the electronic states are quantized into Landau levels, giving rise to chiral edge channels along the sample boundaries \cite{halperin}. Electron waves propagating along these edge channels undergo partial reflection at the QPCs, giving rise to interference. 
The phase difference between the interfering electron waves is determined by the sum of the dynamical and AB phases, $\phi=\phi_{\mathrm{dyn}}+\phi_{\mathrm{AB}}$. The dynamical phase is given by $\phi_{\mathrm{dyn}}=2k_{\mathrm e}d$, where $k_{\mathrm e}$ is the electron Fermi wavevector and $d$ is the effective cavity length. The AB phase is given by $\phi_{\mathrm{AB}}=e\Phi/\hslash$, where $\Phi=BS$ is the magnetic flux enclosed by the area $S$ of the interferometer path.
Following the derivation in Ref.~\cite{setup_pioneer}, assuming equal QPC reflectivities and considering a single interfering edge channel, varying the magnetic field modulates the enclosed magnetic flux, thereby shifting the cavity interference resonances and giving rise to periodic conductance oscillations, described by
\begin{equation}\label{eq:2DEG_eq}
G = \frac{e^2}{h} \left[ 1 + F^* \sin^2\left(\frac{\phi}{2}\right) \right]^{-1},
\end{equation}
where $F^* = \frac{4R^*}{(1-R^*)^2}$ is the electronic coefficient of finesse and $R^*$ is the electronic mirror reflectivity.

To mimic this geometric-phase-induced shift of cavity resonances in optics, we employ the concept of the Spectral Drill cavity.
Its working principle is schematically illustrated in \Cref{fig:wide}a, where a geometric phase shifter (GPS) is placed inside a FP cavity to introduce a geometric phase as light undergoes multiple reflections.
This GPS consists of a fixed quarter-wave plate (QWP1) and a rotatable QWP (QWP2). 
According to Jones matrix calculations (see \Cref{app:derivation}), the total transmittance as a function of the fixed $\mathrm{QWP1}$ orientation angle $\theta_1$ and the rotated $\mathrm{QWP2}$ angle $\theta_2$ can be decomposed in terms of circular cavity eigenmodes as
\begin{align} \label{eq:analytical_drill}
T = {}& \frac{\cos^2(\theta_1)}{1 + F \sin^2(kd + \theta_2 - \theta_1)} \nonumber \\
& + \frac{\sin^2(\theta_1)}{1 + F \sin^2(kd - \theta_2 + \theta_1)},
\end{align}
where $F = \frac{4R}{(1-R)^2}$ is the optical coefficient of finesse, $R$ is the optical mirror reflectivity, $k = \frac{2\pi}{\lambda}$ is the wavevector of the incident light, and $d$ is the cavity mirror separation. 
\Cref{eq:analytical_drill} simplifies to a single Airy distribution for $\theta_1 = 0^\circ$ and $90^\circ$, and yields a symmetric cross-polarization splitting profile for $\theta_1 = 45^\circ$:
\begin{equation} \label{eq:cases}
T = 
\begin{cases} 
\left[ 1 + F\sin^2(kd + \theta_2) \right]^{-1}, & \theta_1 = 0^\circ \\[10pt]
\dfrac{1}{2} \sum_{\pm} \left[ 1 + F\sin^2\left(kd \pm \left(\theta_2 - \frac{\pi}{4}\right)\right) \right]^{-1}, & \theta_{1} = 45^\circ \\[10pt]
\left[ 1 + F\cos^2(kd - \theta_2) \right]^{-1}, & \theta_1 = 90^\circ
\end{cases}
\end{equation}
Consequently, the cavity's axial modes can be controlled through the rotation of $\theta_2$ via the imparted PB phase, which directly emulates \cref{eq:2DEG_eq} of the 2DEG interferometer through a shared mathematical structure combining dynamical and geometric phase contributions.
For completeness, characteristics and similarities between the 2DEG interferometer and our proposed Spectral Drill cavity are summarized in \Cref{tab:optical_vs_electronic_analogy}.

\begin{table*}[tbp]
\caption{\label{tab:optical_vs_electronic_analogy}Analogy between the Spectral Drill cavity and the 2DEG interferometer.}
\begin{ruledtabular}
\resizebox{\textwidth}{!}{%
\begin{tabular}{lll}
\textbf{Feature / Concept} & \textbf{Spectral Drill cavity} & \textbf{2DEG Interferometer \cite{setup_pioneer}} \\  
\hline
Wave type & Electromagnetic wave & Electron wave \\
Geometric phase & Pancharatnam--Berry phase & Aharonov--Bohm phase \\
Parameter space & Polarization space on Poincar\'e sphere & Real space  \\
Control parameter & Rotation angle of QWP ($\theta_2$) & Strength of magnetic field $\perp B$ \\
Phase accumulation & Closed path on Poincar\'e sphere & Closed path in real space \\
Enclosed field & Geometric curvature of Poincar\'e sphere & Magnetic flux \\
Mirror element & Reflecting cavity mirrors & Quantum point contacts \\
Transmittance & $T = \left[ 1 + F\sin^2(\delta/2) \right]^{-1}$ & $G = \frac{e^2}{h} \left[ 1 + F^* \sin^2\left(\phi/2\right)\right]^{-1}$ \\
Phase structure & $\delta /2 = kd + \theta_2 \ (\text{for } \theta_1 = 0^\circ)$ & $\phi/2 = k_e d + e\Phi/2\hslash$ \\
\end{tabular}%
}
\end{ruledtabular}
\end{table*}

Examples of cavity transmittance spectra are shown in \Cref{fig:wide}b, calculated using \cref{eq:cases} for $\theta_1 = 0^\circ$ over a range of incident frequencies and selected rotation angles $\theta_2$. As $\theta_2$ increases, the resonance frequency of a given axial mode decreases to satisfy the resonance condition $kd + \theta_2 = m\pi$.
Consequently, a \ang{180} rotation of QWP2 shifts a given axial mode by one free spectral range (FSR), restoring the initial transmittance spectrum.
More generally, \Cref{fig:wide}c shows the cavity transmittance as a function of mirror displacement at a fixed frequency of 0.3 THz for QWP1 orientation angles $\theta_1$ of $0^\circ$, $45^\circ$, and $90^\circ$.
Similar to \Cref{fig:wide}b, for $\theta_1 = 0^\circ$, the axial modes can be tuned by rotating $\theta_2$. 
The diagonal trajectories of high transmittance correspond to individual axial modes, each defining a line of constant phase.
For $\theta_1 = 90^\circ$, the slope of these diagonal trajectories is reversed, whereas for $\theta_1 = 45^\circ$, modes with both positive and negative slopes appear with equal intensity.
While reversing the slope of the cavity resonances is also achievable in a 2DEG interferometer by reversing the perpendicular magnetic field, the splitting pattern observed for $\theta_1=45^\circ$ arises from both circular polarization components simultaneously acquiring equal and opposite PB phases within the cavity.
To our knowledge, this has no direct counterpart in 2DEG interferometers.

In contrast to conventional mechanisms for tunable FP cavities, such as mechanical mirror displacement \cite{Hercher:68, thz_cavity} or voltage-controlled liquid crystals inside the cavity \cite{Liquid_crystal, Jewell_2008}, our geometric-phase-based approach enables resonance tuning with a fixed mirror geometry and is not subject to limitations such as maximum mirror separation or liquid-crystal drive-voltage saturation. Furthermore, the FSR remains constant during tuning, unlike in conventional methods, where it is accompanied by a slight but exact expansion of the FSR.
\section{\label{sec:exp_setup}Experimental Setup}
To construct the Spectral Drill cavity for operation at the communication frequency of \qty{0.3}{\tera\hertz}, both the QWPs and cavity mirrors must be designed to operate at this frequency.
In this work, we use custom-designed QWPs and mirrors based on subwavelength structures, whose retardance and reflectivity, respectively, are calculated via 3D finite-difference time-domain (FDTD) simulations.
This approach provides rapid prototyping and a degree of customizability not typically available with commercial alternatives, while enabling high-performance THz optical elements that that are easily integrated with resonant tunneling diodes \cite{Saito_2026}.

\subsection{Low-Loss THz Quarter-Wave Plates}
\begin{figure}[tb]
    \centering
    \makebox[\linewidth][l]{(a)}
    \vspace{-6mm}

    \includegraphics[width=0.65\linewidth]{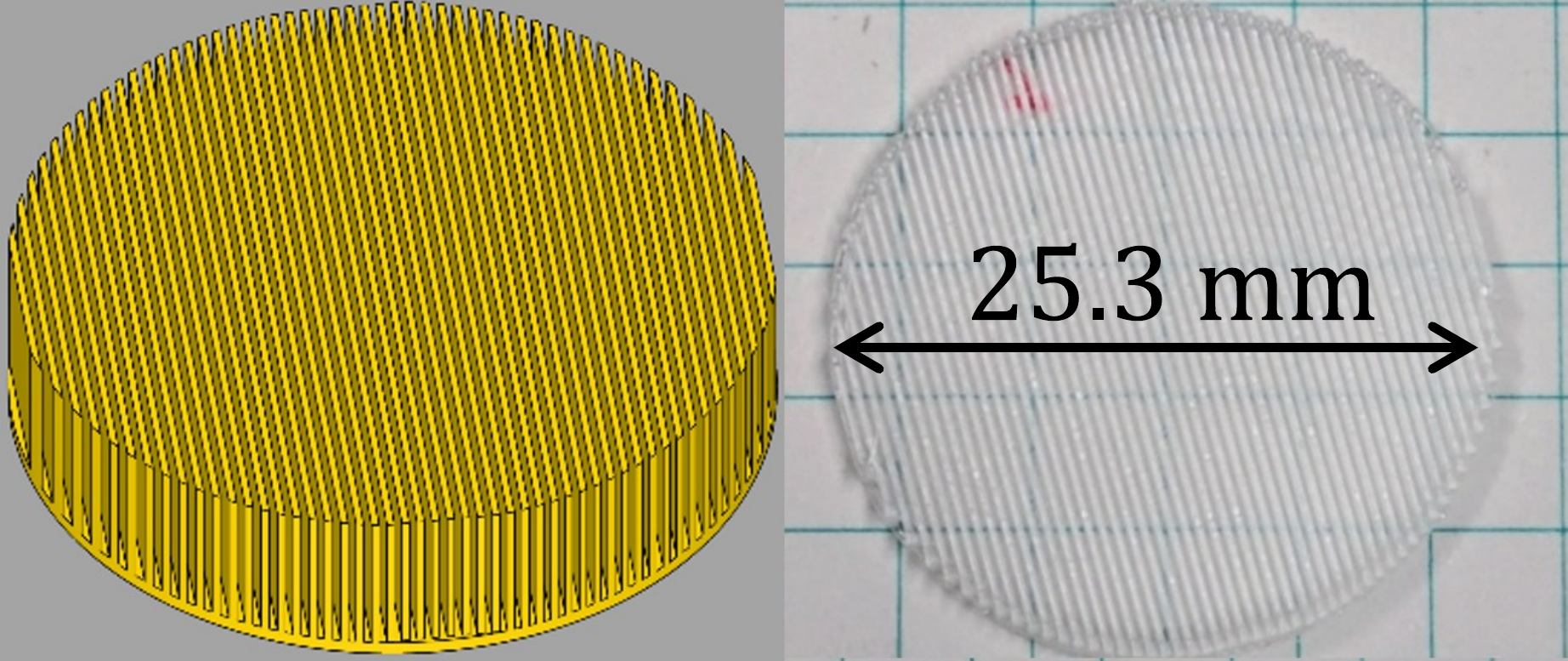}

    \vspace{0mm}

    \makebox[\linewidth][l]{(b)}
    \vspace{-3  mm}

    \includegraphics[
        width=\linewidth,
        trim=0mm 5mm 2mm 5mm,  
        clip
    ]{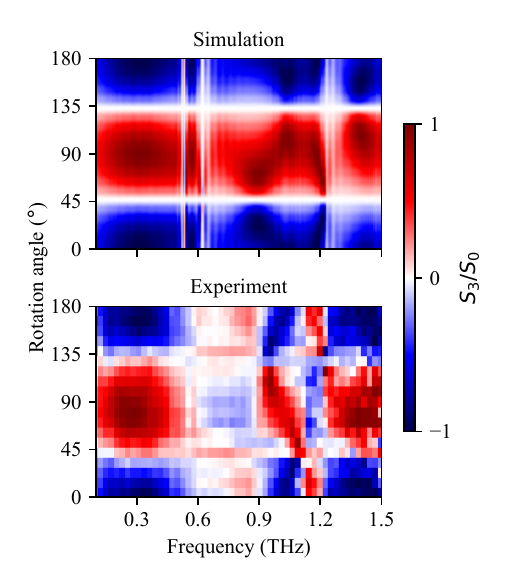}

    \caption{3D-printed QWP: 
    (a) Designed (left) and printed (right) QWP. 
    (b) Polarization spectra obtained via FDTD simulation (top) and THz-TDS experiment (bottom) with rotation.
}
    \label{fig:QWP}
\end{figure}

Intracavity optical losses degrade the cavity finesse and broaden the linewidths of axial modes, making low-loss QWPs essential.
Commercial alternatives, such as quartz waveplates, introduce high reflection losses, while Fresnel rhombs \cite{Kawada:14} suffer from long optical path lengths and high cost.
To overcome these limitations, we utilize custom 3D-printed cyclic olefin copolymer (TOPAS) subwavelength gratings based on our previous work \cite{test}, as shown in \Cref{fig:QWP}a.
Based on form birefringence \cite{rytov1956electromagnetic}, subwavelength gratings can be engineered to function as waveplates by choosing appropriate geometrical dimensions \cite{Rohrbach:21}.
Using FDTD simulations, performed by the open-source Python package PyMeep \cite{OSKOOI2010687}, we determined the grating dimensions to achieve QWP behavior near 0.3 THz, yielding a period of \qty{0.5}{\milli\meter}, a height of \qty{2.65}{\milli\meter}, and a duty cycle of 0.5.
This QWP behavior was confirmed in \cref{fig:QWP}b, where the normalized Stokes parameter $S_3/S_0$, which quantifies the degree of circular polarization, reaches approximately $\mp1$ at rotation angles of \ang{0} and \ang{90}, indicating conversion to left- and right-handed circular polarization, respectively.
Here, the rotation angle measures the angular offset in the counterclockwise direction relative to the \ang{45} angle between the sample and linear input polarization.
The experimental polarization states, measured via THz time-domain spectroscopy (THz-TDS) \cite{Grischkowksy}, are shown at the bottom of \cref{fig:QWP}b, which show excellent agreement with simulations near the frequency of \qty{0.3}{\tera\hertz}.
Above approximately \qty{0.5}{\tera\hertz}, the onset of higher-order diffraction leads to deviations between simulations and experiments, as the finite detector aperture limits the collection of diffracted light.

\subsection{Polarization-Independent THz Cavity Mirrors}
\begin{figure}[t]
    \centering
    \makebox[\linewidth][l]{(a)}
    \vspace{-8mm}

    \includegraphics[width=0.9\linewidth]{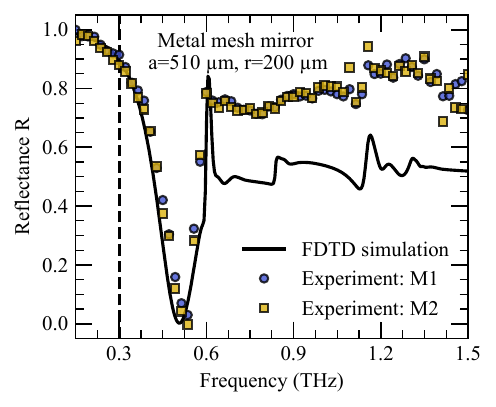}

    \vspace{0mm}

    \makebox[\linewidth][l]{(b)}
    \vspace{-6mm}

    \includegraphics[width=0.9\linewidth]{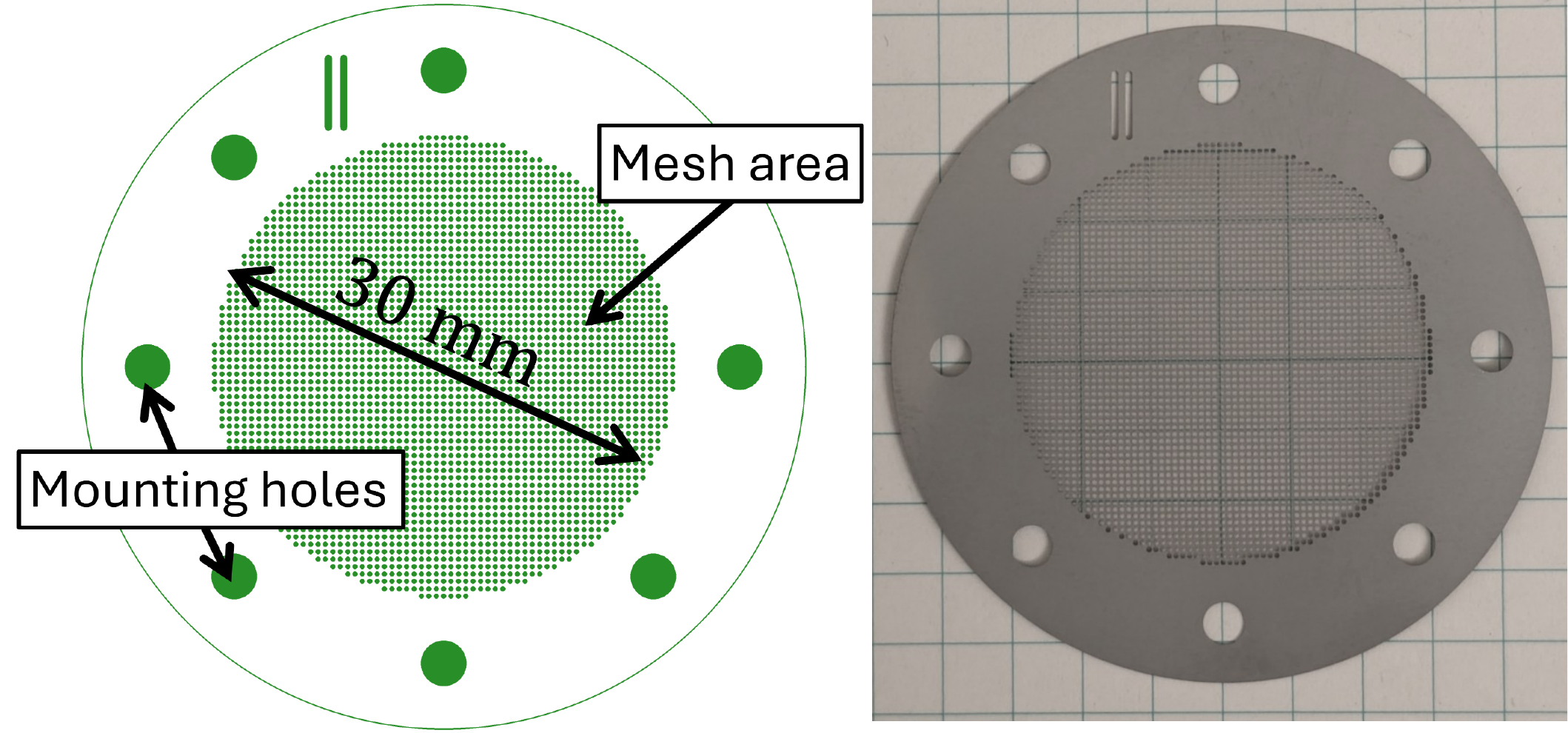}

    \caption{THz cavity mirrors:
    (a) Reflectance spectra obtained from FDTD simulations and THz-TDS measurements of two identical fabricated metal meshes.
    (b) Designed (left) and fabricated (right) metal mesh.}
    \label{fig:mirror_combined}
\end{figure}

At THz frequencies, conventional mirror designs based on dielectric stacks or highly conductive metals such as gold are commonly employed.
However, while these commercial options provide low loss and high reflectivity, their reflectance profile offers limited customizability.
To address this, we design a metal mesh consisting of subwavelength circular holes arranged in a square grid. These meshes serve as customizable THz mirrors, with their reflectance determined by the period $a$ and hole radius $r$.
Importantly, as the polarization state of light inside the Spectral Drill cavity varies with $\theta_2$, polarization-independent mirror reflectance is required to maintain a high cavity finesse. This property is fulfilled by the $C_4$ rotational symmetry of the mesh and was confirmed experimentally.
Based on FDTD simulations (see \Cref{app:mirror}), a design yielding $R = 0.90$ at \qty{0.3}{\tera\hertz} (solid black line in \cref{fig:mirror_combined}a) was chosen, featuring a period of $a = \qty{510}{\micro\meter}$ and a hole radius of $r = \qty{200}{\micro\meter}$.
The reflectance of two fabricated meshes, determined from THz-TDS measurements using $R = 1 - T$, shows good agreement with the simulation results below 0.59 THz, where diffraction effects due to the periodic structure are absent, as shown in \Cref{fig:mirror_combined}a.

\subsection{Assembled Spectral Drill Cavity}
\begin{figure}[t]
    \centering
    \includegraphics[width=0.95\linewidth]{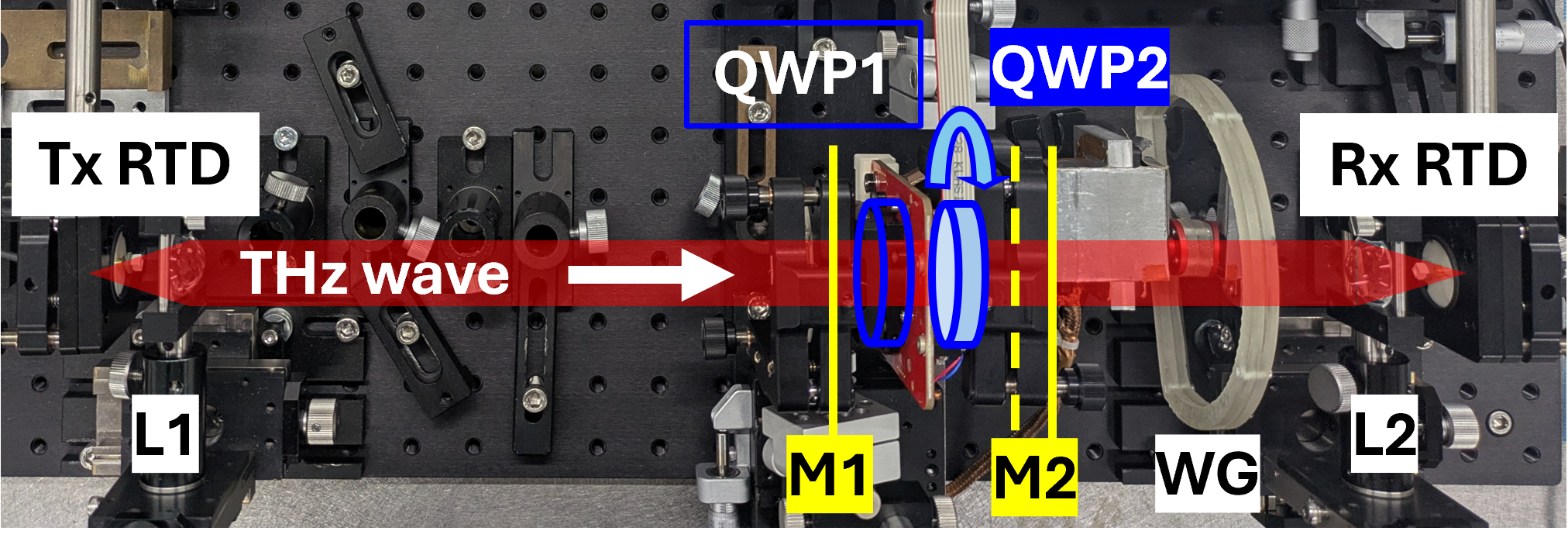}
    \caption{Labeled photograph of the experimental setup. A $\sim\qty{296}{\tera\hertz}$ beam from the transmitter RTD (Tx) is injected into the cavity formed by metal mesh mirrors (M1 and M2) with QWP1 and QWP2 inserted. The transmitted signal is measured at the receiver RTD (Rx) using a wire-grid analyzer (WG) to probe its polarization state. A pair of lenses (L1 and L2) collimate the incoming beam and focus it onto the detector.}
    \label{fig:exp_setup}
\end{figure}

A labeled photograph of the experimental setup is shown in \cref{fig:exp_setup}.
The Spectral Drill cavity consists of two custom-designed, 3D-printed QWPs, marked by the blue cylinders, placed between two custom-designed metal
mesh mirrors, marked by the yellow lines.
THz generation and detection are performed using resonant tunneling diode (RTD) oscillators (ROHM).
A continuous-wave beam at approximately \qty{0.296}{\tera\hertz} is emitted by the transmitter RTD (Tx) and injected into the cavity. The transmitted signal is subsequently measured by the receiver RTD (Rx).
Here, the orientation angle $\theta_1$ of QWP1 is measured relative to the linear polarization angle of the Tx RTD.
The cavity spacing is approximately $L = \qty{45}{\milli\meter}$, leading to a free spectral range of $\text{FSR} = \qty{3.33}{\giga\hertz}$.

During measurements, we record the transmitted signal using under two tuning mechanisms. First, the cavity length is varied to control the dynamical phase. Second, the rotation angle $\theta_2$ of QWP2 is varied to apply the PB phase.

\section{\label{sec:results}Results and Discussion}
\begin{figure*}[t]
  \centering
  \includegraphics[width=0.98\textwidth]{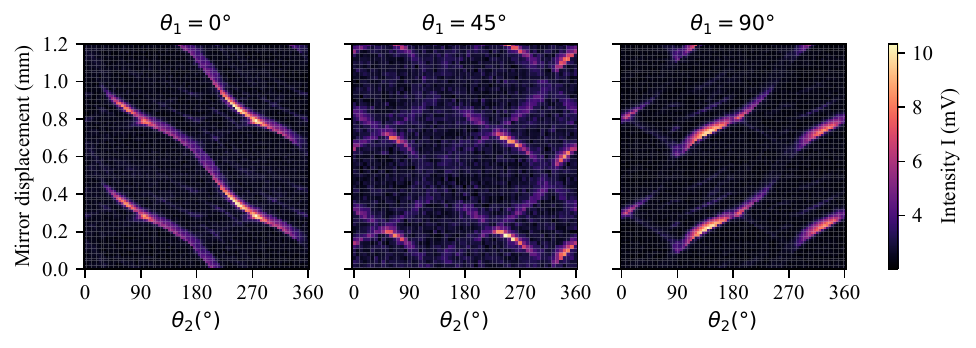}
  \caption{\label{fig:exp_results}
Experimental results of the Spectral Drill cavity for QWP1 orientation angles $\theta_1=\ang{0}, \ang{45},$ and $\ang{90}$.
The transmitted intensity is shown as a function of mirror displacement and QWP2 orientation angle $\theta_2$.
Clear tunability of the axial modes is observed as QWP2 is rotated, with diagonal trajectories of opposite slopes for $\theta_1=\ang{0}$ and $\ang{90}$.
}
\end{figure*}

The experimental results are presented in \cref{fig:exp_results}. For $\theta_1=\ang{0}$, varying the mirror displacement at a constant $\theta_2$ yields high-intensity transmission peaks spaced by approximately \qty{0.5}{\milli\meter}. This spacing corresponds to half the wavelength of the THz source, confirming the observation of neighboring axial modes via conventional dynamical-phase control. Conversely, fixing the mirror displacement while rotating $\theta_2$ continuously shifts the axial modes through the imparted PB phase, demonstrating mode control via the geometric phase of light.
Moreover, the results for other QWP1 orientation angles closely match our analytically derived calculations (\Cref{fig:wide}c). Setting $\theta_1 = \ang{90}$ clearly reproduces the expected slope reversal of the diagonal trajectories, while $\theta_1 = \ang{45}$ confirms the characteristic cross-pattern resulting from coexisting modes with positive and negative slopes.

We observe minor deviations from the analytical calculations:
First, in addition to the fundamental axial modes, fainter diagonal trajectories corresponding to higher-order transverse modes are visible. These higher-order modes, stemming from cavity misalignments, undergo a similar tuning behavior. 
Second, the transmitted intensity fluctuates along the axial mode trajectories, producing regions of high and low intensity due to differential transmission along the fast and slow axes of our 3D-printed QWPs. 
Finally, the trajectories exhibit slight curvatures, which we attribute to the frequency-dependent phase response of our 3D-printed QWPs deviating from ideal quarter-wave retardance.

\begin{figure}[t]
    \centering
    \includegraphics[width=0.9\linewidth]{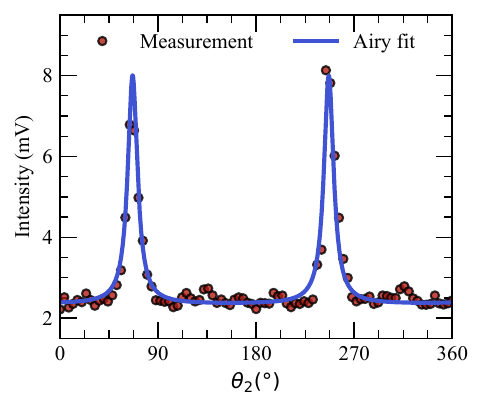}
    \caption{Typical measured spectrum while rotating $\theta_2$, fitted with an Airy distribution.}
    \label{fig:spectrum}
\end{figure}

\Cref{fig:spectrum} shows a typical axial mode spectrum as a function of QWP2 rotation at a fixed cavity length. The transmission profile is characterized by fitting to an Airy distribution of the form
\begin{equation}
I(\theta_2) = I_0 + \frac{A}{1 + F \sin^2(C + \theta_2)},
\end{equation}
in agreement with the analytical model in \cref{eq:analytical_drill} for $\theta_1=0$. 
From the fit parameters, we determine a cavity finesse of $\mathcal{F}=\frac{\pi\sqrt{F}}{2}\approx17$. The corresponding quality factor is given by $Q=m\mathcal{F}\approx1.5\times10^3$, where $m=2L/\lambda\approx90$ is the longitudinal mode order for a cavity length of $L=\qty{45}{\milli\meter}$ at an operating wavelength of approximately $\qty{1}{\milli\meter}$. This corresponds to a FWHM resonance linewidth of $\Delta f_{\mathrm{FWHM}}\approx\qty{200}{\mega\hertz}$, relevant for potential bandpass filtering applications.

\section{\label{sec:conclusion}Conclusion}

In conclusion, we have demonstrated the control of optical cavity modes via the geometric phase of light. Drawing inspiration from 2DEG interferometers tuned by the AB phase, we implemented the first proof-of-concept Spectral Drill cavity at \qty{0.3}{\tera\hertz}. By extending our measurements to include various QWP1 orientation angles $\theta_1$, we verified both the slope reversal of axial mode trajectories at $\theta_1 = \ang{90}$ and the cross-pattern composed of coexisting modes with opposite slopes at $\theta_1 = \ang{45}$, in full agreement with our analytically derived calculations.

To construct our device, we developed custom subwavelength components, including low-loss 3D-printed QWPs and polarization-independent metal mesh mirrors. 
Our cavity achieved a finesse of $\mathcal{F}\approx17$ and a quality factor of $Q\approx1.5\times10^{3}$, yielding a frequency resolution of $\Delta f_{\mathrm{FWHM}}\approx\qty{200}{\mega\hertz}$ with potential applications in bandpass filtering for next-generation wireless communcations.

Finally, the photonic emulation of a 2DEG interferometer demonstrated here extends beyond optics. Because geometric phases appear across different areas of physics, this tuning mechanism can be applied to acoustic or magnetic systems, providing a versatile degree of freedom for wave control.

\begin{acknowledgments}
The authors wish to thank Mr. Kazuisao Tsuruda of ROHM Co., Ltd for
insightful discussions and constructive comments. 

This work was partially supported by Tohoku University-NICT matching fund and JSPS KAKENHI (Grant Number JP25K00930 and JP26K08208).
\end{acknowledgments}

\appendix

\section{Analytical Derivation}\label[appendix]{app:derivation}
Here, we derive the transmission of the Spectral Drill cavity by diagonalizing its round-trip propagation operator, expressed in the circular polarization basis, to identify the cavity eigenmodes and their corresponding resonance conditions.
The transformation between the linear and circular polarization bases is described by the unitary matrices
\begin{equation}
    U = \frac{1}{\sqrt{2}}
    \begin{bmatrix}
    1 & 1\\
    -i & i
    \end{bmatrix}, \quad
    U^{\dagger} = \frac{1}{\sqrt{2}}
    \begin{bmatrix}
    1 & i\\
    1 & -i
    \end{bmatrix}.
\end{equation}
Consequently, diagonally polarized light, represented in the linear basis by the Jones vector $\frac{1}{\sqrt{2}}\begin{bmatrix}1\\1\end{bmatrix}$, is expressed in the circular basis as
\begin{equation}
   E_{\text{in}} =  U^{\dagger}\frac{1}{\sqrt{2}}
    \begin{bmatrix}1\\1\end{bmatrix}
    =
    \frac{1}{2}
    \begin{bmatrix}1+i\\1-i\end{bmatrix}.
\end{equation}
Similarly, the Jones matrix of a rotated QWP, expressed in the circular polarization basis, is given by
\begin{equation}
    J_{\mathrm{circ}}(\theta) =
    \frac{1}{2}
    \begin{bmatrix}
    1+i & (1-i)e^{2i\theta}\\
    (1-i)e^{-2i\theta} & 1+i
    \end{bmatrix}.
\end{equation}
The mirror reflection in the circular basis is described by the matrix
\begin{equation}
    M =
    \begin{bmatrix}
    0 & -1\\
    -1 & 0
    \end{bmatrix}.
\end{equation}

We evaluate the round-trip operator $J_{\text{RT}}$:
\begin{align}
J_{\text{RT}} = {}& M J_{\theta_1}^- J_{\theta_2}^{-}
M J_{\theta_2}^+ J_{\theta_1}^{+} \nonumber\\
= {}& \begin{bmatrix}
\cos\left(2(\theta_2-\theta_1)\right) &
-i e^{-2i\theta_2}\sin\left(2(\theta_2-\theta_1)\right)\\
-i e^{2i\theta_1}\sin\left(2(\theta_2-\theta_1)\right) &
\cos\left(2(\theta_2-\theta_1)\right)
\end{bmatrix}.
\end{align}
where $J_{\theta_2}^+$ and $J_{\theta_1}^+$ represent forward propagation through the QWPs at angles $\theta_1$ and $\theta_2$, respectively, and $J_{\theta_2}^- = J_{-\theta_2}^+$, $J_{\theta_1}^- = J_{-\theta_1}^+$ account for the reversed coordinate system upon reflection. 
Diagonalization of $J_{\mathrm{RT}}$ yields the following diagonal eigenvalue matrix $(\Lambda)$ and normalized eigenvector matrix (V):
\begin{equation}
    \Lambda =
    \begin{bmatrix}
e^{2i(\theta_2-\theta_1)} & 0\\
0 & e^{-2i(\theta_2-\theta_1)}
\end{bmatrix},\quad
V = \frac{1}{\sqrt{2}}
    \begin{bmatrix}
    i e^{2i\theta_1} & -i e^{2i\theta_1}\\
    1 & 1
    \end{bmatrix}.
\end{equation}

The incident field is decomposed into the two cavity eigenmodes by projecting the input Jones vector onto the normalized eigenvectors.
The corresponding mode coupling coefficients are obtained from
\begin{equation}
c_m = v_m^\dagger E_{\text{in}},
\qquad m=1,2,
\end{equation}
where $v_m$ denotes the m-th normalized eigenvector. The fraction of incident power coupled into each eigenmode is 
\begin{equation}
P_m = |c_m|^2.
\end{equation}
The transmission amplitude of each eigenmode is given by 
\begin{equation}
t_m = \frac{1}{1-R e^{2ikd}\lambda_m},
\end{equation}
where $R$ represents the mirror reflectance and $e^{2ikd}$ the round-trip dynamical phase. 
The corresponding intensity transmission is
\begin{equation}
T_m = (1-R)^2 |t_m|^2
= \frac{(1-R)^2}
{|1-R e^{2ikd}\lambda_m|^2}.
\end{equation}
The total cavity transmittance is the sum of the transmitted eigenmodes, given by
\begin{align}
T &= \sum_{m=1}^{2} P_m T_m \nonumber\\
  &= \frac{\cos^2(\theta_1)}
  {1+F\sin^2(kd+\theta_2-\theta_1)}
  \nonumber\\
  &\quad + \frac{\sin^2(\theta_1)}
  {1+F\sin^2(kd-\theta_2+\theta_1)}.
\end{align}

\section{Polarization-Independent THz Cavity Mirrors}
\begin{figure}[b]
    \centering
    \includegraphics[width=0.95\linewidth]{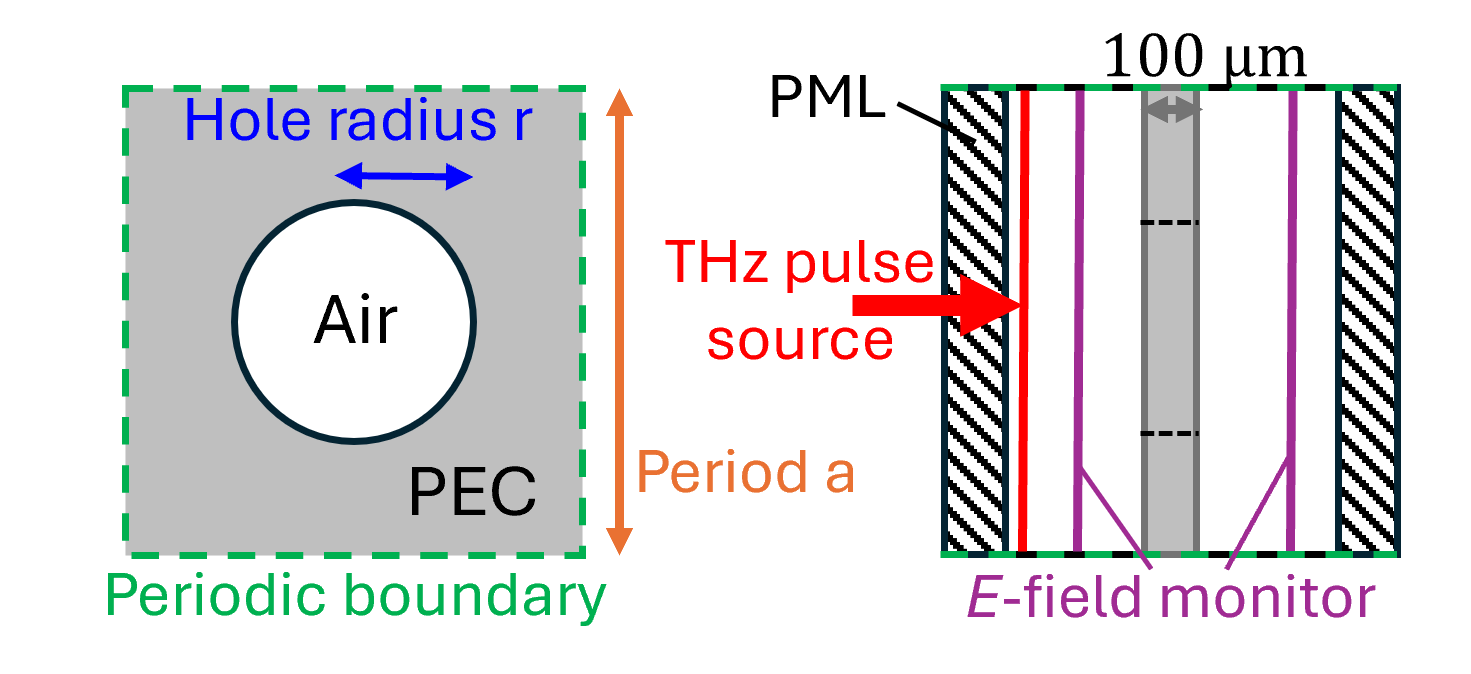}
    \caption{Schematic sketch of the simulation unit cell (left) and the full simulation domain (right) used in the FDTD simulations.}
    \label{fig:sim_domain}
\end{figure}

\label[appendix]{app:mirror}
The details of the FDTD simulations of the proposed metal-mesh THz cavity mirrors are shown in \cref{fig:sim_domain}
The unit cell consists of a \qty{100}{\micro\meter}-thick perfect electric conductor (PEC) sheet with a period $a$, perforated by a central hole of radius $r$.
Periodic boundary conditions are applied at the boundaries of the unit cell. A pulsed THz source is incident normally on the sheet, and the reflectance is derived from the monitored electric fields.

\begin{figure}[!t]
    \centering
    \includegraphics[width=0.98\linewidth]{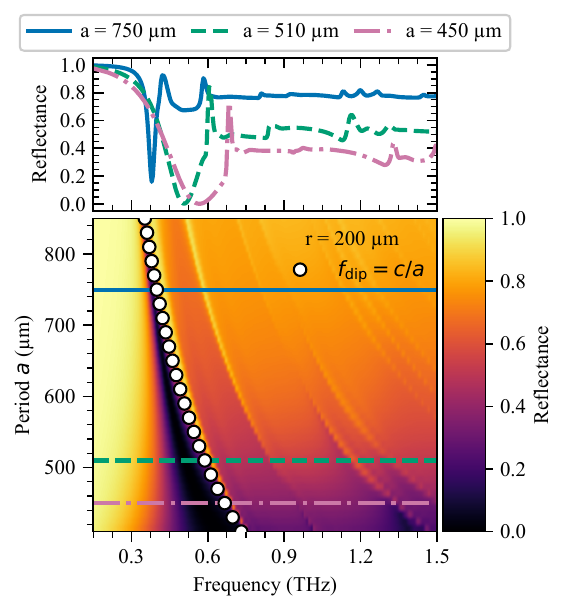}
    \caption{Example FDTD-simulated reflectance spectra of metal meshes for a fixed hole radius of $r=\qty{200}{\micro\meter}$ and varying periods $a$.}
    \label{fig:sim_example}
\end{figure}

Reflectance spectra were simulated for various hole radii and periods. Selected results are shown in \cref{fig:sim_example} for a fixed hole radius of $r=\qty{200}{\micro\meter}$ and varying periods $a$.
The results show resonance dips in the reflectance spectra, with their positions scaling according to $f_{\text{dip}}=c/a$. At this frequency, the first-order diffracted wave transitions from evanescent to propagating.
The initial decrease in reflectance before the resonance dip was used to identify a structure with a target reflectance of $R=0.90$. This was achieved with a period of $a=\qty{510}{\micro\meter}$ and a hole radius of $r=\qty{200}{\micro\meter}$.
Finally, the designed metal-mesh mirrors were laser-cut from stainless steel by a PCB manufacturer.

\bibliography{apssamp}

\end{document}